\documentclass[aps,showpacs,twocolumn]{revtex4}
\usepackage{amsmath}
\usepackage{amsfonts}
\usepackage{amssymb}
\usepackage{graphicx}

\usepackage{indentfirst}

\begin{document}
\title{Recollisions and correlated double ionization with circularly polarized light}

\author{F. Mauger$^1$, C. Chandre$^1$, T. Uzer$^2$}
\affiliation{$^1$ Centre de Physique Th\'eorique, CNRS -- Aix-Marseille Universit\'e, Luminy - case 907, F-13288 Marseille cedex 09, France \\ $^2$ School of Physics, Georgia Institute of Technology, Atlanta, GA 30332-0430, USA}
\date{\today}

\begin{abstract}
It is generally believed that the recollision mechanism of atomic nonsequential double ionization is suppressed in circularly polarized laser fields because the returning electron is unlikely to encounter the core. On the contrary, we find that recollision can and does significantly enhance double ionization, even to the extent of forming a ``knee'', the signature of the nonsequential process. Using a classical model, we explain two apparently contradictory experiments, the absence of a knee for helium and its presence for magnesium. 
\end{abstract}
\maketitle

Multiple ionization of atoms and molecules is usually treated as a rapid sequence of isolated events. However, in the early 90's, experiments using intense laser pulses found double ionization yields which departed from these treatments by several orders of magnitude~\cite{Dorn02}, thereby casting doubt on the uniqueness of the sequential (uncorrelated) multiple ionization channel. These results constitute one of the most striking surprises of recent years in intense laser-matter interactions, and this new ionization channel (``nonsequential double ionization'' (NSDI)~\cite{Dorn02}) has emerged as one of the most dramatic manifestations of electron-electron correlation in nature~\cite{Beck08}. 

Most of the experimental observations of this striking process~\cite{Dorn02} used linearly polarized (LP) laser fields. In this setting, the precise mechanism that makes electron-electron correlation so effective follows the recollision (or ``three-step'') scenario~\cite{cork93,scha93}: An ionized electron, after picking up energy from the field, is hurled back at the ion core upon reversal of the field and dislodges the second electron. NSDI has become an integral part of attosecond physics~\cite{Krau09} since the recollision mechanism requires high-intensity, short-pulse lasers. In fact, it is hoped that attosecond control will provide insights into the dynamics of electron-electron collisions as well as complex multielectron collision phenomena.

Many questions remain unanswered regarding strong-field double ionization, and one that is still completely open concerns polarization. The stakes are high when it comes to understanding the influence of polarization since it is well known that the emission of harmonics in atoms and molecules is strongly dependent on the ellipticity of the driving field~\cite{Budi93,Mair08}, which can therefore act as a control knob. Indeed, recent work shows that elliptic polarization (EP) provides a new control mechanism for recollision physics in high-field ionization~\cite{Wang09}. To illustrate the conceptual difficulty associated with polarization, consider circularly polarized (CP) fields: The recollision scenario which works so well in LP fields is much more difficult to justify in fields in which the ionized electrons tend to spiral out from the core and to miss it~\cite{Brya06}. Therefore one has to come to expect any recollision in CP fields to be due to contamination by small amounts of EP fields, but not otherwise. The matter would rest there if it were not for conflicting experimental evidence: In some experiments using CP fields, the double ionization yields follow the sequential mechanism, confirming current thinking, whereas in others these yields are clearly several orders of magnitude higher than expected, in apparent contradiction with it. The question we resolve here is: Are recollisions possible in pure CP fields or does one have to rely on a small residual ellipticity? 
\begin{figure}
        \includegraphics[width=80mm]{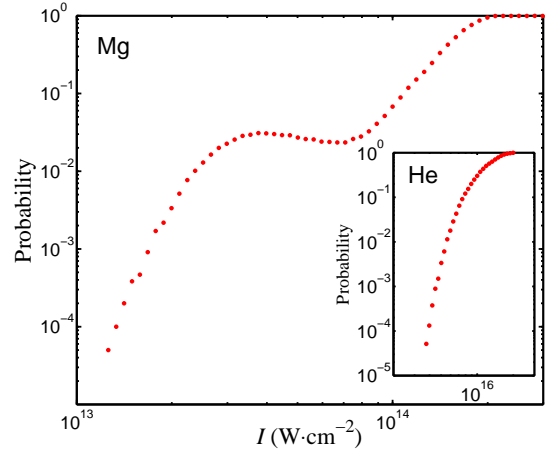}
        \caption {\label{fig1}   
        Computed double ionization probabilites for Hamiltonian~(\ref{Hamiltonian}) as a function of the intensity of the field $I$ for $\omega=0.0584$ a.u. (780 nm) and circular polarization, $a=3$ and $b=1$ (magnesium case). The inset displays the same computation for helium ($a=b=1$).}
\end{figure}

In order to benchmark the process, multiple ionization yields are computed as a function of the intensity of the laser. The signature of the nonsequential double ionization is a characteristic ``knee'' shape in the probability versus intensity curve (see Fig.~\ref{fig1}). Experimentally, a knee has been reported in for nitrogen oxide~\cite{guo01} and for magnesium~\cite{gill01}. Other experiments (including those on helium~\cite{fitt94}) conclude that no knee is formed in CP fields.  

In what follows we explain these seemingly contradictory findings using a classical Hamiltonian model. In addition, we show that, contrary to common belief, recollision can be the dominant mechanism leading to enhanced double ionization yields. The presence or absence of a knee amounts to finding the conditions for the recollision mechanism to work in CP fields. It turns out that enhanced double ionization in CP fields belongs to a class of rotational problems which are closely related to ionization of Rydberg states in CP microwave fields~\cite{farr95} as well as asteroid capture~\cite{genesis} and matter transport~\cite{Koon00} in celestial mechanics. 

The CP field is special among EP fields because viewed from a frame rotating with it, it turns into a constant static field, and the problem becomes analogous to the combined Stark-Zeeman  effect (or the so-called crossed fields problem~\cite{main92,Wang01}) with a substantial Coriolis term which leads to subtleties where the energy is concerned~\cite{Naue90,Howa95}. The key aspect in this frame change is the emergence of the Stark saddle~\cite{Clar85,Howa95} over which pre-ionized electrons penetrate the core or excited core electrons move away from it. Under special conditions the ionized electron can reach the core and  collide with the inner electron, leading to an appreciable amount of NSDI and its characteristic knee signature.

Entirely classical interactions turn out to be adequate to generate the strong two-electron correlation needed for double ionization~\cite{Beck08,ho05_1}. Therefore we work with the classical Hamiltonian model of pseudo-two electron atoms with soft Coulomb potentials~\cite{Java88}. The Hamiltonian is given 
by~\cite{panf01}:
\begin{eqnarray}
    && {\mathcal H}=  \frac{ \left|{\bf p}_{1}\right|^{2}+\left|{\bf p}_{2}\right|^{2}}{2} 
                     -\frac{2}{\sqrt{\left|{\bf x}_{1}\right|^{2}+a^{2}}} 
                     -\frac{2}{\sqrt{\left|{\bf x}_{2}\right|^{2}+a^{2}}} \nonumber \\
    && \quad +\frac{1}{\sqrt{\left|{\bf x}_{1}-{\bf x}_{2}\right|^{2}+b^{2}}}
                     +({\bf x}_1+{\bf x}_2)\cdot {\bf E}(t),                               \label{Hamiltonian}
\end{eqnarray}
where ${\bf x}_i=(x_i,y_i)$ is the position of the $i$-th electron, ${\bf p}_i=(p_{x,i},p_{y,i})$ is its canonically conjugate momentum and $\left|\cdot\right|$ denotes the Euclidean norm in $\mathbb{R}^{2}$. The laser field is circularly polarized, i.e.\ 
${\bf E}(t)= E_{0} f \left( t \right) \left({\bf  e}_x\sin \omega t+{\bf e}_y \cos \omega t\right)$ where $E_0$ is the amplitude of the laser field, $\omega$ its frequency chosen as $\omega=0.0584$~a.u.(780~nm), and
$f \left( t \right)$ the envelope of the pulse with a two laser cycle ramp-up, six laser cycle plateau and two laser cycle ramp down.
The two parameters $a$ and $b$ used in the soft Coulomb potential are determined such that $b$ is sufficiently small to allow significant energy exchange during recollisions, and $a$ such that in the absence of the field, all the initial conditions are bounded (to prevent self-ionization). Therefore the choice of $a$ is related to the binding energy of the ground state: For He suitable parameters are $a=1$ and $b=1$ for a ground state energy ${\mathcal E}_g=-2.24\mbox{ a.u.}$ ~\cite{Haan94}. For Mg, the ground state energy is ${\mathcal E}_g=-0.83~{\rm a.u.}$, and we choose $a=3$ and $b=1$.  We consider a large assembly of initial conditions (typically $1.2\times 10^5$) in the microcanonical ensemble of Hamiltonian~(\ref{Hamiltonian}) in the absence of the field~\cite{Maug09_1}. A statistical analysis of the trajectories provides the double ionization yield versus intensity. The results, reported in Fig.~\ref{fig1}, show a knee for Mg, the characteristic features of which are in very good agreement with  experimental findings~\cite{gill01}~: The correlated double ionization yield peaks at approximately $3\times 10^{13}\ \rm{ W\cdot cm^{-2}}$. A similar calculation performed for He (inset of Fig.~\ref{fig1}) does not give any evidence of a knee, once again in agreement with experiment~\cite{fitt94}.
 
\begin{figure}
        \includegraphics[width=80mm]{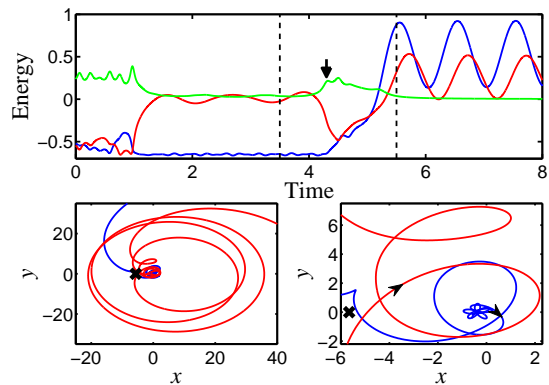}
        \caption {\label{fig2}   
        NSDI in magnesium with a laser intensity $I=2\times10^{13} \ \mbox{W}\cdot\mbox{cm}^{-2}$. The upper panel displays the energy (kinetic energy plus Coulomb interaction with the nucleus) of each electron (red and blue curves) as a function of time and the green curve represents the energy of the (Coulomb) repulsion between the two electrons. The arrow indicates the moment of recollision. Lower left panel: The position of each electron in the rotating frame. Lower right panel: The position of each electron in the rotating frame during the time interval indicated by the dashed lines in the upper panel. The arrows indicate the positions of the electrons at the recollision and the direction of their motion. In the lower panels we display the position of the saddle point by a cross.}
\end{figure}  

For Mg, with a CP laser field intensity $I=2 \times 10^{13}~\mbox{W}\cdot\mbox{cm}^{-2}$ it turns out that approximately $6\%$ of the trajectories are subjected to a strong recollision (i.e., an electron is ionized and then returns to approach the other one closer than 3.2 a.u.). Thus, in this case, counterintuitively, recollisions are not marginal and do play a leading role in the observed knee. In what follows, we use tools from nonlinear dynamics~\cite{Maug09_1} to identify the mechanism by which the electrons undergo correlated double ionization. In a nutshell, the correlated double ionizing trajectories follow a modified three-step model: First, an electron is ionized by escaping through a saddle. If it is trapped close to a periodic orbit (this family of outer-electron periodic orbits will be explained in the next section), it might return to the core through the saddle, recollide with the core electron, and result in correlated double ionization when both electrons escape, once again, through the saddle. In Fig.~\ref{fig2} we display a typical example of such a trajectory.
 
We have noted before~\cite{Maug09_1} that, in the absence of the field, a typical two-electron trajectory is composed of one electron close to the nucleus (the ``inner'' electron) and another further away (the ``outer'' electron), with quick exchanges of the roles of each electron. This distinction is crucial when the laser field is turned on~: The outer electron is mainly driven by the field, whereas the field competes with the nuclear interaction in driving the inner electron. Therefore we can use reduced models to identify the route the outer electron follows to the core. As in the LP case~\cite{Maug09_1}, the cooperation of the inner and outer electrons is essential for ionization in CP.


{\em Outer electron dynamics--}  
As for LP, finding recolliding trajectories amounts to finding the initial conditions ${\bf x}_{0}$ and ${\bf p}_{0}$ in phase space such that the trajectory returns to its initial position (in the core region) after some recollision time $t_{\rm recoll}$, i.e., by solving the system ${\bf x}\left( t_{\rm recoll} \right) = {\bf x}_{0}$. A reduced model for the outer electron dynamics is obtained by neglecting the interaction with the nucleus and with the other electron~\cite{Maug09_1}. In this case, the maximum recolliding kinetic energy brought back by the outer electron is $\kappa U_{p}$ where $\kappa \approx 3.17$~\cite{cork93,scha93} (and $U_p$ is the ponderomotive energy) as it is exactly the case for LP. The analysis of recolliding trajectories shows that this is not the dominant behavior observed for the outer electron and therefore is not the main mechanism for recollisions. The correct mechanism emerges from viewing the reduced model in a rotating frame, where we compute the distance $\rho(t)$ between the position of the electron and $(E_0/\omega^2,0)$ as
$$
\rho (t)=\Vert \tilde{\bf x}_0+\tilde{\bf p}_0 t\Vert,
$$
where $\tilde{\bf x}_0={\bf x}_0-(E_0/\omega^2,0)$ and $\tilde{\bf p}_0={\bf p}_0-(0,E_0/\omega)$. 
For $\tilde{\bf p}_0=0$, the outer electron follows a circular periodic orbit (with constant momentum) and so stays in the vicinity of the core region. This is actually what is seen from a typical recolliding trajectory: $\tilde{\bf p}_{0} \approx 0$ so that it is located in the neighborhood of one of the periodic orbits (with $\rho(t)$ constant, see lower left panel of Fig.~\ref{fig2}). 
The condition $\tilde{\mathbf{p}}_{0} \approx 0$ gives a natural criterion for the feasibility of recollisions: If the condition corresponds to admissible initial conditions, recollisions are possible. In contrast, if the condition $\tilde{\mathbf{p}}_{0} \approx 0$ is far away from admissible initial conditions, no recolliding trajectories are possible. For 780 nm and $I = 2 \times 10^{13} \ \mbox{W} \cdot \mbox{cm}^{-2}$ the recollision condition gives $\mathbf{p}_{0} \approx 0.4$ while for $I = 10^{15}$, it gives $\mathbf{p}_{0} \approx 2.9$. The first case corresponds to admissible conditions and thus explains why we observe recollisions in Mg. In contrast, the second case is not allowed which explains the absence of recolliding trajectories for He. 

\begin{figure}
        \includegraphics[width=80mm]{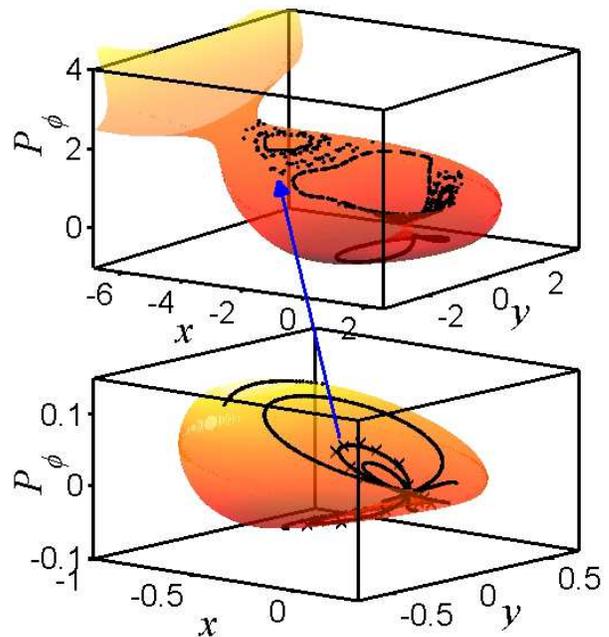}
        \caption {\label{fig3}   
        Poincar\'e sections of the inner electron reduced model in the 3D space $\left( x,y,P_{\phi}=xp_{y}-yp_{x} \right)$. The admissible set is represented by the surface and Poincar\'e sections are represented by dots. Lower panel: Sections for a Jacobi constant ${\mathcal K}=-0.659$ which corresponds to the constant for the inner electron in Fig.~\ref{fig2} when it is close to the nucleus and the outer electron is far away. For comparison, we display the Poincar\'e section associated with this part of the trajectory for the full model (crosses). Upper panel: Poincar\'e surface for a Jacobi constant ${\mathcal K}=-0.5$, when the ionization channel is open. The arrow represents the jump, from an invariant torus up to the chaotic unbounded area, experienced by the inner electron during NSDI.}
\end{figure}

Starting close to the circular periodic orbit, the outer electron finds its way through the saddle point to the core region because of the Coulomb interaction with the nucleus, and might collide with the other electron. Depending on the efficiency of the collision~\cite{Maug10_1}, both electrons may leave the core, resulting in NSDI (see lower right panel of Fig.~\ref{fig2}).
A model for the outer electron dynamics which takes into account the interaction with the nucleus is obtained from Hamiltonian~(\ref{Hamiltonian}) by neglecting the interaction with the other electron. This dynamics is best appreciated in the rotating frame where the reduced Hamiltonian becomes
\begin{equation}
 {\mathcal K}=\frac{\left|{\bf p}\right|^{2}}{2}-\frac{2}{\sqrt{\left|{\bf x}\right|^{2}+a^{2}}}
              -\omega ( xp_y-yp_x) +E_0 x,
\label{HamIN}
\end{equation}
where ${\bf x}=(x,y)$ is the position of the electron in the rotating frame, ${\bf p}=(p_x,p_y)$ its canonically conjugate momentum, and ${\mathcal K}$ is the Jacobi constant~\cite{Hill78}. 
The saddle point is located at $x=x_*$, $y=0$, $p_x=0$ and $p_y=\omega x_*$ where $x_*$ is a real (negative) solution of $\omega^2 x-E_0-2x/(x^2+a^{2})^{3/2}=0$ which corresponds to a saddle of the so-called zero-velocity surface~\cite{Hill78}. When the ionization channel is open (depending on the Jacobi constant, see Fig.~\ref{fig3}), it becomes possible for an electron to penetrate the core region. This is exactly what happens to the outer electron which is in the vicinity of the circular periodic orbits. The trajectory displayed in Fig.~\ref{fig2} (lower left panel) shows that the outer electron enters the core region close to the saddle indicated by a cross.


{\em Inner electron dynamics--} Hamiltonian~(\ref{HamIN}) is also a model for the dynamics of the inner electron, but it is only valid in the core region where the effect of the potential is strong so that the inner electron is bound. In order to investigate its dynamics, we construct Poincar\'e sections of Hamiltonian~(\ref{HamIN}) with a surface of equation $xp_x+yp_y=0$ (see Fig.~\ref{fig3}). For the relevant range of Jacobi constants, the dynamics shows an elliptic island in the core region which binds the inner electron. At some critical Jacobi constant an ionization channel opens up, raising the possibility for the inner electron to leave this bound region after a small exchange of energy during a recollision with the outer electron. The ionization channel through a saddle is similar to the one observed for the ionization of Rydberg atoms in strong CP microwave fields through electron collisions with the core~\cite{farr95}. 
The behavior of the inner electron is determined by its Jacobi constant: For small Jacobi constants, the accessible set is composed of two disconnected surfaces, a bounded one close to the core and another, unbounded one which corresponds to ionized positions. Because of the choice of initial conditions for the atom -- on the ground state energy surface -- this second component is initially not accessible to the inner electron. However, for larger Jacobi constants, the two surfaces merge into a single one, thus establishing a connection between the core and unbound regions.

When the Jacobi constant is small, the set which is accessible to the inner electron is organized by invariant tori which fill up all the accessible domain of the inner electron (see Fig.~\ref{fig3}, lower panel). But when the constant increases, some invariant tori break up, leading to coexistence between regular structures and a chaotic sea (see upper panel, Fig.~\ref{fig3}). This chaotic sea, connected to the unbound region, defines an ionization channel for the inner electron. This channel passes close to a saddle point which gives a natural guide for the ionization seen in the rotating frame. The way for the inner electron to move from an invariant torus associated with a small Jacobi constant up to the chaotic sea is through collisions with the outer electron when it returns to the core region. This jump in the Jacobi constant is symbolized by an arrow in Fig.~\ref{fig3}.
The higher the Jacobi constant is, the larger is the channel of ionization around the saddle point. Thus, when the channel is very narrow, the inner electron may be trapped for a while before finding its way out and ionize, leading to a delay between the last collision and ionization of the inner electron. The amount of energy the inner electron gains from the returning electron determines its chances to ionize by leaving an invariant torus to reach the chaotic sea where it is swept away by the laser field (see Fig.~\ref{fig3}). Recollisions lead to double ionization if the inner electron gains enough energy to jump into the chaotic (unbound) region while leaving the outer electron with enough energy to remain ionized. Such special recollisions are the main source of enhancement of the double ionization probability with a circularly polarized field and the formation of a knee with varying intensity.

We are grateful to P. Ranitovic for enlightening discussions. CC acknowledges financial support from the PICS program of the CNRS. This work is partially funded by NSF.



\end{document}